\newif\ifprint

\printtrue

\documentclass[class/draft]{my-class/my-class}

\usepackage{my-goodcharter}
\usepackage{microtype}

\usepackage{mathtools,amssymb}

\usepackage{bm}

\usepackage{my-semantic-logic-symbols}

\usepackage[
  shortcuts,
  allowbreakbefore,
]{extdash}

\usepackage[
  bibstyle=alphabetic,
  citestyle=alphabetic,
  backend=biber,
  natbib=true,
]{biblatex}
\usepackage[inline]{enumitem}

\usepackage{hyperref}
\usepackage[open,openlevel=2]{bookmark}

\usepackage{orcidlink}
\usepackage[
  mycolor=print,
]{my-cactus}

\usepackage{my-thm}

\mathchardef\mhyphen="2D
\DeclareMathOperator{\pr}{pr}
\DeclareMathOperator{\Obs}{O}
\DeclareMathOperator{\Con}{C}
\DeclareMathOperator{\Dx}{D}
\mytitle{
  Equivalence of Decentralized Observation, Diagnosis, and Control Problems in Discrete\-/event Systems
}
\mydate{\today}

\myauthor{K.~Ritsuka}
\myIEEEmembership{Student Member}
\myorcid{0000-0003-3713-5964}
\myemail{Ritsuka314@queensu.ca}
\myaddressref{qu}

\myauthor{K.~Rudie}
\myIEEEmembership{Fellow}
\myorcid{0000-0002-8675-334X}
\myemail{karen.rudie@queensu.ca}
\myaddressref{qu}

\myaddress{qu}{
  Department of Electrical and Computer Engineering and Ingenuity Labs Research Institute\\
  Queen's University, Kingston, ON, Canada K7L 3N6
}

\mykeywords{
  Automata, Discrete event systems, Supervisory control, Agents and autonomous systems, Cooperative control
}

\myabstract{
  This paper demonstrates an equivalence between
  observation problems,
  control problems (with partial observation),
  and diagnosis problems of
  decentralized discrete-event systems, namely,
  the three classes of problems are Turing equivalent,
  as one class Turing reduces to another.
  
  The equivalence allows decomposition of a control problem into a collection of simpler control sub\-/problems,
  which are each equivalent to an observation problem;
  and similarly allows converting a diagnosis problem to a formally simpler observation problem.
  Since observation problems in their most general formulation have been shown to be undecidable in previous work,
  the equivalence produced here demonstrates that control problems
  are also undecidable;
  whereas the undecidability of diagnosis problems
  is a known result.
}

\begin{document}

\mymaketitle

\section{Introduction}
\label{sec:intro}

Most research in discrete\-/event systems (DES)
falls into two categories:
those concerning closed\-/loop systems
such as control problems,
and
those concerning open\-/loop systems,
such as observation problems and diagnosis problems.

\begin{figure*}
  \centering
  \includegraphics[width=0.8\textwidth]{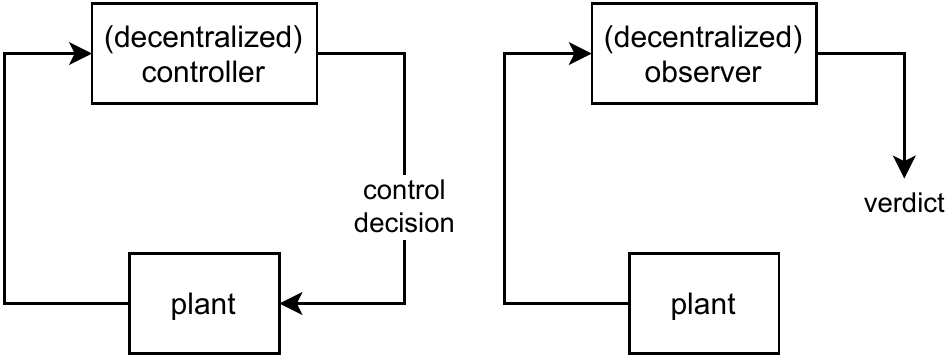}
  \caption[Distinction between open\-/loop systems and close\-/loop systems]{
    Left: open\-/loop systems. Right: closed\-/loop systems. Adapted from \citep{Tripakis_VLTS21}.}
  \label{fig:problem}
\end{figure*}

For a discrete\-/event system plant,
a closed\-/loop system is formed
by imposing supervisory control over the plant.
A control problem asks for a supervisory control policy
so that the closed\-/loop system meets some prescribed properties.
The scheme of control problems is illustrated in \cref{fig:problem}.

Studies of control problems
began with the seminal work of \citet{Ramadge1987}.
Partial observations \citep{Lin1988d} and
decentralized supervision \citep{Cieslak1988,Rudie1992}
were introduced in subsequent studies.
\Citet{Cieslak1988} and \citet{Rudie1992} initially introduced decentralized supervision
under a constraint of available local control decisions
and how overall control decisions are fused from the local ones.
That constraint has been gradually relaxed
\citep{Prosser1997,Yoo2002,Yoo2004,Kumar2005,Chakib2011}
over the past few decades.

On the other hand,
open\-/loop systems take different forms.
A concrete example is diagnosis problems \citep{Sampath1995,Debouk2000,Sengupta2002,Qiu2006,Wang2007}
that seek distinguishing strings contain ``faulty'' events
within a bounded delay of the occurrence of the faulty events.
On the other hand,
a more abstract example is observation problems
that seek distinguishing strings from a prescribed collection.
The earliest formalization of observation problems,
as known to the author,
is by \citet{Tripakis2004}.
The scheme of open\-/loop systems is illustrated in \cref{fig:problem}.

The three classes of problems,
control, diagnosis, and observation,
seem to be unrelated.
Control problems concern closed\-/loop behaviour,
whereas diagnosis problems allow a delay for the correct verdicts to be made,
but observation problems do not.
Therefore,
the three classes of problems are usually studied separately.

However,
results for one of the classes of problems
have often been adopted
to one of the other classes of problems.
This suggests that there is a mutual connection
between the three classes of problems.
This document is intended to establish such connection
as an equivalence between the three classes of problems.

\section{Observation Problem}

An observation problem seeks to distinguish strings in a set $K$,
where $K \subseteq L$,
from strings in $L - K$.
Formally, an observation problem is specified as follows.
Given alphabet $\Sigma$ and subalphabets $\Sigma_{i,o} \subseteq \Sigma$ called the observed alphabets,
natural projections $P_i\colon \Sigma^* \to \Sigma_{i,o}$,
for agents $i \in \nset = \set{1, \dots, n}$,
and given languages $K\subseteq L \subseteq \Sigma^*$,
the observation problem is to construct observers
$f_i$ and a fusion rule $f$,
such that
\begin{equation}
  \label{eq:observation}
  \begin{aligned}
    & \All{s \in L} 
    \\
    & \quad
      \begin{alignedat}[t]{2}
        & s \in K     && \implies f(f_1P_1(s), \dots, f_nP_n(s)) = 1
        \\ {}\land{}
        & s \in L - K && \implies f(f_1P_1(s), \dots, f_nP_n(s)) = 0
      \end{alignedat}
  \end{aligned}
\end{equation}

An instance of the observation problem, $\mathsf{Obs}$, is denoted by
$\Obs(L, K, \setidx{\Sigma_{i,o}}{i \in \nset})$
or more simply,
$\Obs(L, K, \Sigma_{i,o})$.

If the fusion rule $f$ is given as part of the problem,
then the instance is denoted by $\Obs(f, L, K, \Sigma_{i,o})$.
Such problems are instance of the $f$-observation problem,
or $f\mhyphen\mathsf{Obs}$.

Solvability of observation problems is known to be undecidable \citep{Tripakis2004}.

We will show that diagnosis problems and control problems
are both equivalent to observation problems.

\section{Diagnosis Problem}

The diagnosis problems
were first studied in the centralized case by \citet{Sampath1995},
and extended to the decentralized cases \citep{Debouk2000,Sengupta2002,Qiu2006,Wang2007}.

A diagnosis problem seeks to
identify strings containing special events,
known as ``faulty events'',
within a bounded delay of time.
Formally, a diagnosis problem is specified as follows.
Given alphabet $\Sigma$ and subalphabets $\Sigma_{i,o} \subseteq \Sigma$ called the observed alphabets,
natural projections $P_i\colon \Sigma^* \to \Sigma_{i,o}$,
for agents $i \in \nset = \set{1, \dots, n}$.
For a fault alphabet $\Sigma_f \subseteq \Sigma_{uo} = \Sigma - \Union_i \Sigma_{io}$,
a language $L$,
say a string $s \in L$ is positive (faulty) if $s$ contains at least one symbol from $\Sigma_f$,
and otherwise is negative.
We may assume that there is a single fault event $\sigma_f$
as this assumption is inconsequential to the hardness of the problem.

For a faulty string $s$, if $s = \pi \sigma_f \tau$ for some strings $\pi$ and $\tau$,
where $|\tau| \ge m$,
we say that $s$ is faulty for at least $m$ steps.
In other words, a string $st$ is faulty for at least $m$ steps
if $s$ is faulty and $|t| \ge m$.

Then the diagnosis problem is,
given an upper bound of delay as an integer $m$,
construct observers $f_i$ and a fusion rule $f$,
such that
\begin{equation}
  \label{eq:diagnosis}
  \begin{aligned}
    & \All{s \in L} 
    \\
    & \quad
      \begin{alignedat}[t]{2}
        & \text{$s$ is positive for at least $m$ steps} && \implies f(f_1P_1(s), \dots, f_nP_n(s)) = 1
        \\ {}\land{}
        & \text{$s$ is negative}                        && \implies f(f_1P_1(s), \dots, f_nP_n(s)) = 0
      \end{alignedat}
  \end{aligned}
\end{equation}
That is, faulty strings are diagnosed after at most $m$ steps of the fault.

The problem statement above is a simplification:
A subtlety in the problem statement is that there may exist a faulty string $s \in L$
that is positive for less than $m$ steps but
has no extension in $L$ which is positive for at least $m$ steps.
Since we nonetheless want to diagnose such faulty strings,
the phrase ``$s$ is positive for at least $m$ steps''
should be augmented to include such strings.

An instance of diagnosis problem, $\mathsf{Dx}$, is denoted by
$\Dx(L, \setidx{\Sigma_{i,o}}{i \in \nset}, \sigma_f, m)$
or more simply,
$\Dx(L, \Sigma_{i,o}, \sigma_f, m)$.

If the fusion rule $f$ is given as part of the problem,
then the instance is denoted by $\Dx(f, L, \Sigma_{i,o}, \sigma_f, m)$.
Such problems are instance of the $f$-diagnosis problem,
or $f\mhyphen\mathsf{Dx}$.

\subsection{Equivalence of Diagnosis Problems and Observation Problems}

\begin{thm}
  \label{thm:dx-to-obs}
  Given a fusion rule $f$,
  the class of $f$-diagnosis problems\---$f\mhyphen\mathsf{Dx}$\---
  reduces to 
  the class of $f$-observation problems\---$f\mhyphen\mathsf{Obs}$.
\end{thm}

\begin{pf}
  For a given $f$-diagnosis problem $\Dx(f, L, \Sigma_{i,o}, \sigma_f, m)$,
  construct the following $f$-observation problem $\Obs(f, L, K, \Sigma_{i,o})$,
  where $K = \setcomp{s \in L}{\text{$s$ is positive for}\allowbreak\text{at least $m$ steps}}$.

  By construction, \cref{eq:diagnosis,eq:observation} coincide.
\end{pf}

\begin{thm}
  \label{thm:obs-to-dx}
  Given a fusion rule $f$,
  $f\mhyphen\mathsf{Obs}$
  reduces to 
  $f\mhyphen\mathsf{Dx}$.
\end{thm}

\begin{pf}
  For a given $f$-observation problem $\Obs(f, L, K, \Sigma_{i,o})$,
  construct the following $f$-diagnosis problem $\Dx(f, L', \Sigma_{i,o}, \sigma_f, 0)$,
  where
  $L' = (L-K) \union \setcomp{s\sigma_f}{s \in K}$
  and 
  where we have chosen $m = 0$.

  Notice that negative strings in $L'$
  are exactly strings in $L - K$,
  and a string in $L'$ that is positive (for at least $0$ steps)
  \--- i.e., one in $\setcomp{s\sigma_f}{s \in K}$ \---
  uniquely corresponds to a string $s \in K$ and
  satisfies $s' = s\sigma_f$,
  hence $P_i(s) = P_i(s'\sigma_f^n) = P_i(s')$.
  Thus,
  by construction, \cref{eq:observation,eq:diagnosis} coincide.
\end{pf}

\begin{thm}
  The classes of problems
  $f\mhyphen\mathsf{Obs}$ and $f\mhyphen\mathsf{Dx}$
  are equivalent.
  Moreover,
  $\mathsf{Obs}$ and $\mathsf{Dx}$
  are equivalent.
\end{thm}

\begin{pf}
  By \cref{thm:dx-to-obs,thm:obs-to-dx}.
\end{pf}

It is known that solvability of diagnosis problems is undecidable \citep{Sengupta2002}.
The reduction \cref{thm:dx-to-obs} offers an alternative route to proving that undecidability.
Namely, we showed that observation problems reduce to
diagnosis problems, and from \citet{Tripakis2004}
we know that observation problems are undecidable.

\section{Control Problem}

Recall that
the control problem is to construct controllers
$f^\sigma_i$ and fusion rules $f^\sigma$,
for each event $\sigma \in \Sigma_c$,
such that
\begin{equation}
  \label{eq:control}
  \begin{aligned}
    & \All{s \in K}
    \\
    & \quad
      \begin{alignedat}[t]{2}
        & s\sigma \in K     && \implies f^\sigma(f^\sigma_1P_1(s), \dots, f^\sigma_{n_\sigma}P_{n_\sigma}(s)) = 1
        \\ {}\land{}
        & s\sigma \in L - K && \implies f^\sigma(f^\sigma_1P_1(s), \dots, f^\sigma_{n_\sigma}P_{n_\sigma}(s)) = 0
      \end{alignedat}
  \end{aligned}
\end{equation}

To avoid trivial unsolvable instances,
we assume that an instance is always controllable.

An instance of control problem, $\mathrm{Con}$, is denoted by
$\Con(L,\allowbreak K,\allowbreak \setidx{\Sigma_{i, o}}{i \in \nset},\allowbreak \setidx{\Sigma_{i, c}}{i \in \nset})$,
or more simply,
$\Con(L,\allowbreak K,\allowbreak \Sigma_{i, o},\allowbreak \Sigma_{i, c})$.

\subsection{Equivalence of Control Problems and Observation Problems}

We first revise the problem specification of the control problems.

\begin{thm}
  Define the following two languages
  \begin{equation}
    \label{eq:LK}
    \begin{aligned}
      L_\sigma &= \setcomp{s \in K}{s\sigma \in L} \\
      K_\sigma &= \setcomp{s \in K}{s\sigma \in K}.
    \end{aligned}
  \end{equation}
  Then \cref{eq:control} is equivalent to
  \begin{equation}
    \label{eq:control-2}
    \begin{aligned}
      & \All{\sigma \in \Sigma_{c}, s \in L_\sigma}
      \\
      & \quad
        \begin{alignedat}[t]{2}
          & s \in K_\sigma            && \implies f(f^\sigma_1P_1(s), \dots, f^\sigma_{n_\sigma}P_{n_\sigma}(s)) = 1
          \\ {}\land{}
          & s \in L_\sigma - K_\sigma && \implies f(f^\sigma_1P_1(s), \dots, f^\sigma_{n_\sigma}P_{n_\sigma}(s)) = 0.
        \end{alignedat}
    \end{aligned}
  \end{equation}
\end{thm}

\begin{pf}
  By definition,
  for all $s \in L_\sigma$,
  \[
    s\sigma \in L - K \iff s \in L_\sigma - K_\sigma.
  \]
  This concludes the proof.
\end{pf}

\begin{thm}
  \label{thm:con-to-obs}
  The classes of problems $\mathsf{Con}$ reduces to $\mathsf{Obs}$
\end{thm}

\begin{pf}
  For a given control problem $\Con(L, K, \Sigma_{i,o}, \Sigma_{i,c})$,
  construct the following observation problems
  $$\setidx{\Obs(L_\sigma, K_\sigma, \setidx{\Sigma_{i,o}}{i \in \nset_\sigma})}{\sigma \in \Sigma_c}.$$
  
  By construction, \cref{eq:observation,eq:control-2} coincide.
\end{pf}

From the proof we can see that
it is appropriate to decompose a control problem
into a collection of individual (control) sub-problems,
each one dealing with a specific event.

\begin{thm}
  \label{thm:obs-to-con}
  The class of problems $\mathsf{Obs}$ reduces to $\mathsf{Con'}$.
\end{thm}

\begin{pf}
  For a given observation problem $\Obs(L, K, \Sigma_{i,o})$,
  construct a control problem as follows.
  First add to the alphabet a distinguished letter $\gamma$,
  and let $\Sigma_{i,c} = \set{\gamma}$ for all $i \in \nset$.
  Henceforth, let $\pr(M)$ stands for the prefix\-/closure of language $M$.
  Now let
  \begin{alignat*}{3}
    &L' &{}:={}& \pr(L\gamma) \\
    &   &{} ={}& \pr(L) \union L\gamma \\
    &K' &{}:={}& \pr(K\gamma \union L) \\
    &   &{} ={}& \pr(K) \union K\gamma \union \pr(L). \\
  \end{alignat*}
  The control problem is then
  \[\Con(L', K', \Sigma_{i,o}, \Sigma_{i,c}).\]
  It should be verified that the control problem is well\-/posed.
  First, it is clear that $L'$ and $K'$ are indeed prefix\-/closed.
  To verify controllability, let $\Sigma$ be the alphabet of $L$,
  and hence $\Sigma_{uc} = \Sigma$.
  Then, for any $\sigma \in \Sigma_{uc}$ and string $s \in K'$, 
  suppose that $s\sigma \in L'$.
  If $s\sigma \in \pr(L)$,
  $s\sigma \in K'$ as desired.
  If $s\sigma \in L\gamma$,
  then $\sigma = \gamma$,
  which contradicts the fact that $\gamma$ is a controllable event.

  Now compute the languages in \cref{eq:LK}.
  First, we have
  \begin{alignat*}{5}
    L'_\gamma &= \setcomp{s \in K' &&}{s\gamma &&\in L'                   &&} \\
              &= \setcomp{s \in K' &&}{s\gamma &&\in \pr(L) \union L\gamma &&} \\
              &= \setcomp{s \in K' &&}{s       &&\in L                    &&} \\
              &= L
  \end{alignat*}
  where the third line is due to
  $\gamma$ being a distinguished letter that is not in $L$,
  and consequently not in $\pr(L)$;
  the fourth line is due to the facts that
  $L \subseteq K\gamma \union L \subseteq \pr(K\gamma \union L) = K'$.
  Similarly, we have
  \begin{alignat*}{5}
    K'_\gamma &= \setcomp{s \in K' &&}{s\gamma &&\in K'                                &&} \\
              &= \setcomp{s \in K' &&}{s\gamma &&\in \pr(K) \union K\gamma \union \pr(L) &&} \\
              &= \setcomp{s \in K' &&}{s       &&\in K                                 &&} \\
              &= K
  \end{alignat*}
  where the third line is due to $\gamma$ being a distinguished letter that is not in $L$,
  and also $K$ being a subset of $L$.
  The last line is due to $K \subseteq \pr(K) \subseteq K'$.
  Then \cref{eq:control-2} coincides with \cref{eq:observation}.
\end{pf}

\begin{thm}
  \label{thm:Obs=Con}
  The classes of problems $\mathsf{Obs}$ and $\mathsf{Con}$
  are equivalent.
\end{thm}

\begin{pf}
  By \cref{thm:con-to-obs,thm:obs-to-con}.
\end{pf}

The approach of \citet{Lin1988d}
in dealing with centralized control problems under partial observation
can be interpreted as a special case of the reduction of control problems
to observation problems (i.e., $\mathrm{CON} \leq_T \mathrm{OBS}$).

\begin{corr}
  \label{corr:con-undecidable}
  Solvability of control problems are undecidable in general.
\end{corr}

\begin{pf}
  We have just shown that the observation problems reduces to control problems,
  whereas \citeauthor{Tripakis2004} demonstrated that
  solvability of observation problems is undecidable \citep{Tripakis2004}.
\end{pf}

\Cref{corr:con-undecidable}
only states the undecidability of control problems
when \emph{no restriction} is placed on the fusion rule.
However,
in special cases when the fusion rule is restricted,
such as for the architecture given by \citet{Cieslak1988,Rudie1992},
solvability can still be decided \citep{Rudie1995}.

\printbibliography[segment=\therefsegment, heading=subbibintoc]

\end{document}